\documentclass[showpacs,preprintnumbers,amsmath,amssymb,twocolumn,superscriptaddress]{revtex4}
\usepackage{graphicx}
\usepackage{dcolumn}
\usepackage{bm}
\usepackage{ulem}

\begin{document}
\newcommand{\eq}{{\,=\,}}
\newcommand{\duu}{\nabla^{\left\langle\mu\right.}u^{\left.\nu\right\rangle}}
\newcommand{\begq}{\begin{equation}}
\newcommand{\eneq}{\end{equation}}
\newcommand{\bqa}{\begin{eqnarray}}
\newcommand{\eqa}{\end{eqnarray}}

\setlength{\abovedisplayskip}{3pt plus1pt minus1pt}
\setlength{\belowdisplayskip}{3pt plus1pt minus1pt}
\setlength{\textfloatsep}{10pt plus1pt minus1pt}
\setlength{\abovecaptionskip}{0pt}

\draft

\title{Effects of bulk viscosity on hadron spectra and  Hanbury-Brown
Twiss radius by the causal viscous hydrodynamics
}\thanks{Supported in part by the National Natural Science
Foundation of China  under Grant No 10610285, 10775167 and
10705044, and  the Knowledge Innovation Project of Chinese Academy
of Sciences under Grant No. KJCX2-YW-A14 and KJCX3-SYW-N2, the
Startup Foundation for the CAS Presidential Scholarship Award
under Grant No. 29010702.}

\author{J. W. Li}
\affiliation{Shanghai Institute of Applied Physics, Chinese
Academy of Sciences, P.O. Box 800-204, Shanghai 201800, China}
\affiliation{Graduate School of the Chinese Academy of Sciences,
Beijing 100080, China}

\author{ Y. G.  Ma} \thanks{Corresponding author. E-mail: ygma@sinap.ac.cn}
\affiliation{Shanghai Institute of Applied Physics, Chinese
Academy of Sciences, P.O. Box 800-204, Shanghai 201800, China}
\author{ G. L.  Ma}
\affiliation{Shanghai Institute of Applied Physics, Chinese
Academy of Sciences, P.O. Box 800-204, Shanghai 201800, China}

\date{\today}

\begin{abstract}
The particle spectra and Hanbury-Brown Twiss (HBT) radius of Au+Au
collisions  at RHIC energy are investigated by a hydrodynamical
expanding source with both shear and bulk viscosities ($\zeta$).
With a large width of the ratio
 of $\zeta$ to entropy density $s$, both the particle
transverse momentum spectra and the ratio $R_{out}/R_{side}$ of
HBT radii in the direction of the total transverse momentum of
detected two particles ($R_{out}$) and perpendicular to both this
direction and the beam direction ($R_{side}$) become a little
steeper.
\end{abstract}



 \maketitle

\section{Introduction}
Ideal hydrodynamics has successfully explained the experimental
data of elliptic flow coefficient $v_2(p_T)$ as well as the
transverse momentum spectra of hadrons in central and
semi-peripheral collisions up to $p_T$ = 1.5--2\,GeV/$c$ and also
predicted the hard photon production.$^{[1,2]}$ However the almost
perfect ideal fluid dynamical description of the experimental data
gradually breaks down in the regions of more peripheral collisions
and/or high transverse momenta ($p_T$), and it also fails to
reproduce the HBT radius.$^{[3,4]}$ Actually ideal fluid is only
an approximation,$^{[5]}$ people need to investigate the viscous
hydrodynamics. The so-called the first order relativistic viscous
hydrodynamics was firstly formulated by Eckart and a variation by
Landau and Lifshitz, and many calculations have been done so
far.$^{[6-8]}$ But there were two problems in their approaches:
one is the dissipative fluctuations which may propagate at a speed
larger than the speed of light and thus lead to the causality
problem; the other is the solutions which may develop
instabilities. The second order formalism which was developed
about thirty years ago by Israel and Stewart avoided these
difficulties.$^{[9]}$ But due to the complication of the evolution
equations, numerical codes have not been developed to solve the
Israel-Stewart equations until recent years.$^{[10-16]}$ When
compared to the experimental data, people only considered the
shear viscosity and found that the particle spectra and elliptic
flow became more consistent in larger $p_T$ range. But the HBT
radius are still not consistent with the experimental
data.$^{[12]}$

Although one expects that the bulk viscosity is to be negligible
for temperatures far away from the phase transition temperature
$T_c$, large effects are expected near $T_c$.$^{[17-21]}$ As
suggested in \cite{Pratt}, the HBT results may come close to the
experimental data if bulk viscosity is taken into account. Based
on this argument, we want to check if this argument is valid in
the present work. To this end, we use the algorithm which has been
used in \cite{Roma} to consider the effect of the bulk viscosity
for central Au+Au collision at RHIC energies.

\section{Causal viscous hydrodynamics and results}
The general hydrodynamic equations arise from the local
conservation of energy and momentum$^{[1]}$
\begin{eqnarray}
\label{t}
  \partial_\mu T^{\mu \nu}(x)&=&0,
\end{eqnarray}
where the energy-momentum tensor without heat conduction is
decomposed in the form$^{[14]}$
\begin{eqnarray}
\label{tmunu}
  T^{\mu\nu}&=& eu^{\mu}u^{\nu} - (p{+}\Pi)\Delta^{\mu \nu}
               + \pi^{\mu \nu}.
\end{eqnarray}
Here $e$ and $p$ are the local energy density and thermal
equilibrium pressure, and $u^{\mu}$ is the 4-velocity of the
energy flow which obeys $u^\mu u_\mu = 1$. $\Pi$ is the bulk
viscous pressure, $\Delta^{\mu\nu} = g^{\mu\nu}-u^{\mu}u^{\nu}$ is
transverse to the flow velocity, that is $\Delta^{\mu\nu}u_{\nu} =
0$. $\pi^{\mu \nu}$ is the traceless shear viscous pressure
tensor. With Eq.~(\ref{t}) and Eq.~(\ref{tmunu}), we can get the
evolution equations of the energy density and the 4-velocity of
the energy flow \bqa
&&\!\!\!\!\!\!\!\!\!\!\!\!\!\!\!\!\!\!\!\!\!\!\!\!\!\!\!\!\!
(\epsilon+p+\Pi)D u^\mu = \nabla^\mu (p+\Pi)
\nonumber\\
&&~~~~~~~~~~~-\Delta^\mu_{\nu} \nabla_\sigma \pi^{\nu
\sigma}+\pi^{\mu \nu} D u_\nu\, , \label{2.1} \eqa \bqa D \epsilon
&=& - (\epsilon+p+\Pi) \nabla_\mu u^\mu+\pi^{\mu \nu}
\langle\nabla_\nu u_\mu\rangle\, , \label{2.2} \eqa where $D$ is
the time derivative in the local fluid rest frame and fulfills $D
= u^{\mu}\partial_{\mu}$. The angular bracket notation is defined
by $\langle\nabla^\mu u^\nu\rangle = \frac{1}{2}(\nabla^\mu
u^\nu{+}\nabla^\nu u^\mu) -
\frac{1}{3}(\nabla{\cdot}u)\Delta^{\mu\nu}$. In the Israel-Stewart
approach  the kinetic evolution equations of the bulk pressure
$\Pi$ and the traceless shear viscous tensor $\pi^{\mu\nu}$ are
$^{[14]}$
\begin{eqnarray}
\label{Pi-transport}
  D{\Pi}&=&-\frac{1}{\tau_{\Pi}}\big(\Pi+\zeta \nabla{\cdot}u\big),
\\
  D\pi^{\mu \nu} &=&-\frac{1}{\tau_{\pi}}\big(\pi^{\mu\nu}-2\eta\duu\big)
\nonumber\\
\label{pi-transport}
  && -\bigl(u^\mu\pi^{\nu\alpha} + u^\nu\pi^{\mu\alpha}\bigr)
      Du_\alpha,
\end{eqnarray}
where $\eta$ and $\zeta$ denotes the bulk and shear viscous
coefficients, respectively. $\tau_{\pi}$ and $\tau_{\Pi}$ are the
relaxation times for the bulk pressure and the shear viscous
pressure tensor, respectively. They can be related to $\eta$ and
$\zeta$ as follows $^{[10]}$
\begin{eqnarray}
\label{rel}
\tau_{\pi}=2\eta\beta_2~~,~~~~~~~~~~~\tau_{\Pi}=\zeta\beta_0.
\end{eqnarray}

For a massless Boltzmann gas $\beta_2 \approx \frac{3}{4p}$, so we
can get $\tau_{\pi}\approx \frac{3\eta}{2p}$ which translates to
$\tau_{\pi}\approx \frac{\eta}{s}\frac{6}{T}$,$^{[10,11]}$ where
$s$ is entropy density and this is the value we use in this paper.
In this case the relaxation and coupling coefficients for bulk
pressure vanish, $\beta_0\to \infty$ and thus there are no effects
of the bulk viscosity.$^{[10]}$ But as we already know the bulk
viscous coefficient $\zeta$ becomes large near the quark-hadron
phase transition temperature $T_c$.$^{[23]}$ So the behavior of
the bulk viscous pressure must change and $\beta_0$ does not tend
to be infinite any more. However, as to our knowledge, there are
no quantitative results for $\beta_0$ that are relevant for
applications to nuclear collisions (an attempt has been made in
\cite{Muronga}). In this paper we will try two different $\beta_0$
in our calculations which correspond to two different
$\tau_{\Pi}$, namely, $\tau_{\Pi} = \frac{\zeta}{s}\frac{8}{T}$
and $\tau_{\Pi} = \frac{\zeta}{s}\frac{30}{T}$, respectively. Of
course people could try smaller $\beta_0$. But with a smaller
$\beta_0$, there will be casuality problem and this may arise
instability.$^{[24]}$

As in ideal hydrodynamics we need initial conditions at $\tau_0$,
equation of state of the bulk matter and freeze-out condition to
obtain the results of these evolution equations. Generally there
are two models for the initialization of the hydrodynamics:
Glauber-type model and Color Glass Condensate-type model. Here we
use the Glauber model and the energy density at the initial time
$\tau_0$ is parameterized by the number density of wounded
nucleons $\epsilon \sim n_{\rm WN}$,$^{[12,25]}$ where $n_{\rm WN}
= 2T_A[1-(1-\frac{\sigma T_A}{A})^A]$ and $T_A =
\int_{-\infty}^{\infty}{\rm d}z\rho_0/\{1+{\rm
exp}[(\sqrt{r^2+z^2}-R_0)/\chi]\}$. For gold nuclei, $ A = 197$,
$R_0 = 6.4$ fm, $\chi = 0.54$ fm, the nucleon-nucleon cross
section $\sigma$ at $\sqrt{s_{NN}} = 200$ GeV is assumed to be
$40$ mb and $\rho_0$ is chosen to fulfill $2\pi\int r {\rm d}rT_A
= A$. The EOS we use here is calculated by Laine and Schroder
which corresponds to a phase transition temperature $T_c\approx
190 \pm 20$ MeV.$^{[26]}$ In their calculation they considered a
hadron resonance gas at low temperatures and used high order
weak-coupling QCD result at high temperatures. And the freeze-out
mechanism is the Cooper-Frye prescription that is particles freeze
out at a single temperature $T_f$.$^{[27]}$

\begin{center}
{\includegraphics*[width=0.34\textwidth]{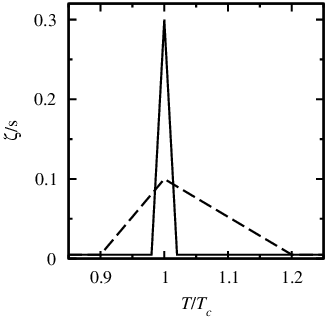}}
\begin{minipage} {0.44\textwidth}
\footnotesize {\bf Fig.1.}\ {\footnotesize Two linearly changed
$\zeta/s$ with different width and height. See texts for details.}
\end{minipage}
\end{center}

 Unlike ideal hydrodynamics, in causal viscous
hydrodynamics we need to give the initial condition for
$\pi^{\mu\nu}$ and $\Pi$ as well as the values of $\eta/s$ and
$\zeta/s$.$^{[10-12,14]}$ Here we choose $\pi^{\mu\nu} = 0$ and
$\Pi = 0$ at the initial time which minimize the effects of
viscosity. There have been many calculations for $\eta/s$  in
which there is a conjectured minimal bound for $\eta/s$ in AdS/CFT
that $\eta/s \ge 1/4\pi \approx 0.08$,$^{[28]}$ and we can see in
\cite{Xuzhe} that $\eta/s$ does not change a lot for different
temperatures. In this paper we want to see the influence of the
bulk viscosity, so we use the lower bound of $\eta/s$ during the
whole evolution of the fluid. In \cite{Li} the value of $\zeta/s$
reaches the maximum at $T_c$ and changes linearly to a very small
value when the temperature goes away from $T_c$ and the width is
small.However, in \cite{Muller,Meyer} the width is large. In this
paper, we will take two linearly changing $\zeta/s$ with different
width and height as shown in Fig.1 in our calculations.

For central Au+Au collisions at RHIC, the geometries which are
longitudinally expanding are space-time rapidity independent and
have radial symmetry. For these geometries it is convenient to
work in the co-moving and radial coordinates $\tau,r,\phi,\eta$
with the relations $\tau = \sqrt{t^2-z^2}$, $r^2 = x^2+y^2$,
$\tan{\phi} = y/x$ and $\eta = {\rm atanh} (z/t)$.

The only non-vanishing fluid velocity components are then $u^\tau$
and $u^r$ with the relation $u^\tau = \sqrt{1+(u^r)^2}$. To solve
Israel-Stewart hydrodynamic equations which have this kind of
geometry we use the one dimensional algorithm used in \cite{Roma}.

As the procedure that adopted in the ideal hydrodynamics,$^{[30]}$
the initial central temperature $T_0$ and the freeze-out
temperature $T_f$ are chosen to make both the normalization and
the slope of the resulting pion spectrum in reasonable agreement
with the experimental data and we get $T_0  =  0.34$ MeV and $T_f
= 0.17$ MeV. The freeze-out temperature is larger than that
typically used in ideal hydrodynamics. In ideal hydrodynamics
interactions of the system can keep the system in thermal
equilibrium, but the interactions are not so efficient in viscous
hydrodynamics which allows for departures from equilibrium. So an
earlier freeze-out and thus a higher $T_f$ is to be expected in
viscous hydrodynamics. In \cite{Roma2} there are other parameters
that can fit both the pion spectrum and the HBT radii, but either
the $\eta/s$ is too large or the freeze-out temperature is
unreasonable large.

Fig.2 shows the velocity and temperature profiles for different
times of the evolution of the fluid using the parameters after
fitting the pion spectra. The left and right panels are for
relaxation time values $\tau_{\Pi} = \frac{\zeta}{s}\frac{8}{T}$
and $\tau_{\Pi} = \frac{\zeta}{s}\frac{30}{T}$, respectively. In
each panel, we calculate three cases: (a) $\eta/s = 0.08$ and
without bulk viscosity, (b) $\eta/s = 0.08$,
$(\frac{\zeta}{s})_{max} = 0.3$ and the later changes linearly to
$(\frac{\zeta}{s})_{min} = 0.005$ at $0.98T_c$ and $1.02T_c$
respectively, corresponding to the solid line in Fig.1, and (c)
$\eta/s = 0.08$, $(\frac{\zeta}{s})_{max} = 0.1$ and the later
changes linearly to $(\frac{\zeta}{s})_{min} = 0.005$ at $0.9T_c$
and $1.2T_c$ respectively, corresponding to the dashed line in
Fig.1. We can see that if bulk viscosity is taken into account, at
earlier time the velocity changes near $T_c$, and in case (c) the
velocity changes more rapidly than case (b). This is due to the
narrower width and larger height of $\zeta/s$. And with a much
larger peak there may be non-periodic oscillation in velocity and
temperature. So the method of hydrodynamics may not be suitable
and the source may clusterize to small  fragments.$^{[31]}$ At
later time, the effect of bulk viscosity extends to other place
where the temperature is far away $T_c$. We can see that the
evolution of the fluid is sensitive to the width of $\zeta/s$
rather than its height. In both case (b) and (c), the fluid tends
to be retarded and the effect is stronger with larger width. We
can also see that the effect of bulk viscosity is  a little
stronger with a smaller relaxation time.

\begin{center}
{\includegraphics*[width = 0.46\textwidth]{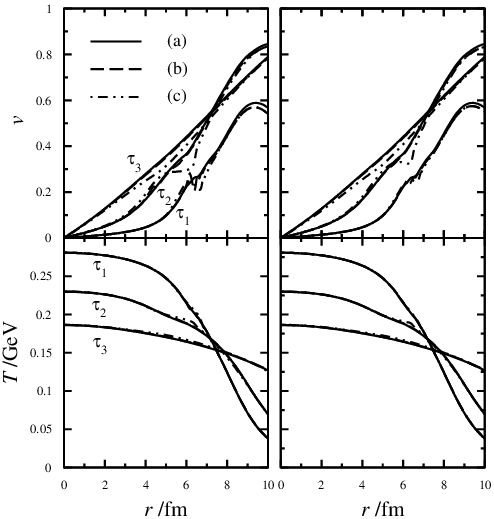}}
\begin{minipage}
{0.48\textwidth}\footnotesize {\bf Fig.2.}\ {\footnotesize
Temperature and velocity profiles for different conditions (a),
(b) and (c) with $\tau_{\Pi}  =  \frac{\zeta}{s}\frac{8}{T}$ (left
panels) and $\tau_{\Pi}  =  \frac{\zeta}{s}\frac{30}{T}$ (right
panels) of the relaxation time values at different times $\tau_n
= 2n $fm/c. See texts for details.}
\end{minipage}
\end{center}

We plot the pion, kaon and proton spectra together with the RHIC
STAR and PHENIX experimental data for the most central $5\%$ of Au
+ Au collisions at $\sqrt{s_{\rm NN}} = 200$ GeV in
Fig.3.$^{[32,33]}$ It shows that case (a) and (b) are almost the
same and case (c) become steeper which indicates smaller
transverse flow. From this we can conclude that the effect of the
width is more important than the height of $\zeta/s$ in
hydrodynamics.

\begin{center}
{\includegraphics*[width = 0.3\textwidth]{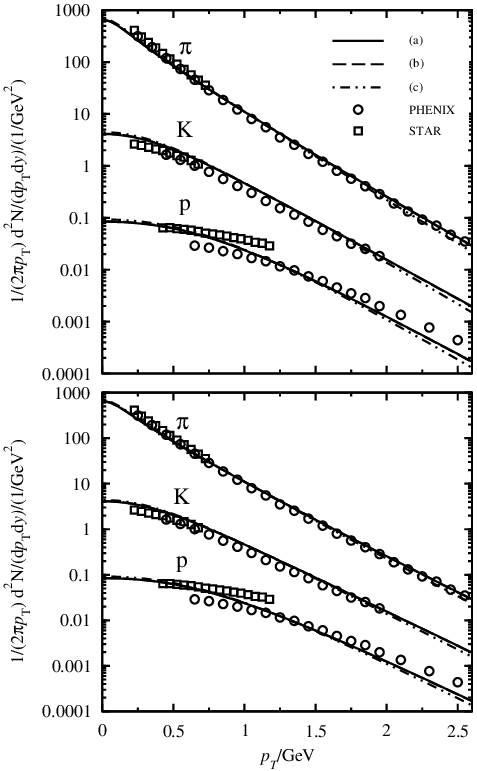}}
\begin{minipage}{0.46\textwidth}\footnotesize {\bf Fig.3.}\ {\footnotesize
Comparison of the calculated spectra of $\pi$, $K$ (scaled by 0.1)
and $P$ (scaled by 0.01) for different conditions (a), (b) and (c)
with the STAR and PHENIX data. The upper and bottom panel
corresponds to $\tau_{\Pi} = \frac{\zeta}{s}\frac{8}{T}$ and
$\tau_{\Pi} = \frac{\zeta}{s}\frac{30}{T}$ of the relaxation time
values, respectively.}
\end{minipage}
\end{center}

Finally we also show the viscous hydrodynamic results for the HBT
radii together with experimental data. The results are shown in
Fig.4 where the ratios $R_{out}/R_{side}$ are for different
relaxation time values and different cases.$^{[34]}$ The radius
$R_{out}$ represents the radius in the direction of the total
transverse momentum of detected two particles and $R_{side}$
corresponds to the radius perpendicular to both the total
transverse momentum direction and the beam direction. There are
some quantitative effects for HBT radius' ratio in different $p_T$
range. Essentially $R_{out}/R_{side}$ becomes steeper versus the
transverse momentum when the bulk viscosity is taken into account,
which is consistent with the steeper transverse momentum spectra
observed in Fig.3. Nevertheless we still did not see the bulk
viscosity can change the ratio substantially. This also can be
seen in \cite{Pratt2}. The reason could be that the
thermodynamical quantities change only a little at $T_f$ for
different cases, and the freeze-out surface is almost the same.

\section{Summary and discussions}
In summary, we have solved the causal viscous hydrodynamic
equations for central Au+Au collisions at
 $\sqrt{s_{\rm NN}} = 200$
GeV using a simple numerical code.
 Using the parameters that fitted
the pion spectra, we
\begin{center}
{\includegraphics*[width = 0.27\textwidth]{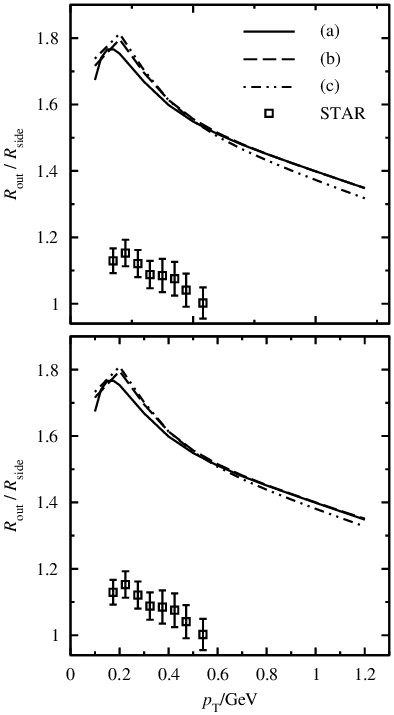}} \vspace{0.2cm}
\begin{minipage}{0.46\textwidth}
\footnotesize {\bf Fig.4.}\ {\footnotesize Calculated ratio
$R_{out}/R_{side}$ of HBT radii for different conditions (a), (b)
and (c) together with STAR data. The upper and bottom panel
corresponds to $\tau_{\Pi} = \frac{\zeta}{s}\frac{8}{T}$ and
$\tau_{\Pi} = \frac{\zeta}{s}\frac{30}{T}$ of the relaxation time
values, respectively.}
\end{minipage}
\end{center}
calculated the temperature and velocity
 profiles, the pion, kaon and proton spectra as well
as the HBT results for different relaxation time values and
different bulk viscosity cases. We find that the fluid evolves
slowly when the bulk viscosity is taken into account. The particle
spectra become steeper which indicates smaller transverse flow
when the width of $\zeta/s$ becomes larger. We can also see that
the HBT results are still far away from the experimental data when
the bulk viscosity is taken into account in our approach.

There are other problems that have to be considered to obtain a
proper description of the experimental HBT radii. Firstly, we
should investigate the influence of different EOS. In \cite{Pratt}
the author use different EOS to calculate the causal viscous
hydrodynamics and find the behavior of $R_{out}/R_{side}$ for
different $\zeta/s$ changes substantially. Secondly, larger
$\zeta/s$ may develop instabilities in the mixed phase, and the
source may be clusterized to droplets and this freeze-out
mechanism is different from Cooper-Frye formalism.$^{[31,36]}$ A
granular source method was claimed that it can explain the HBT
puzzle.$^{[37]}$\\\\\\
 {\bf \Large Acknowledgment}\\

We thank Dr. P. Romatschke and Dr. Mei Huang for communications.

\end{document}